\documentclass[intlimits,twoside,a4paper]{article}

\usepackage{amsmath,amssymb}
\usepackage{graphicx}
\newcommand*\pct{\scalebox{.9}{\%}}
\usepackage{color}
\usepackage[T2A]{fontenc}
\usepackage[cp1251]{inputenc}

\usepackage[eqsecnum]{cmpj3}

\issue{2017}{20}{3}{33602}
\doinumber{10.5488/CMP.20.33602}

\title[On the improvement of SPT2 approach]%
{On the improvement of SPT2 approach in the theory of a hard sphere fluid in disordered porous media} %
\author[M. Holovko, T. Patsahan, W. Dong]{M. Holovko\refaddr{label1}, T. Patsahan\refaddr{label1}, W.~Dong\refaddr{label2}}

\addresses{
\addr{label1} Institute for Condensed Matter Physics of the
National Academy of Sciences of Ukraine, \\1 Svientsitskii St.,
79011 Lviv, Ukraine
\addr{label2} \'Ecole Normale Sup\'erieure de Lyon, Laboratoire de Chimie, UMR 5182 CNRS,\\ 46 All\'ee d'Italie, 69364 Lyon, Cedex 07, France }

\date{Received July 7, 2017, in final form August 5, 2017}

\begin{document}
\maketitle

\begin{abstract}
The SPT2 approach is based on the scaled particle theory and developed for the description
of thermodynamic properties of hard sphere (HS) fluids in disordered porous media.
Using this approach a porous medium is modelled as a quenched matrix of hard spheres (HS) or overlapping
hard spheres (OHS). A hard sphere fluid immersed in a matrix can move in a void between matrix particles.
A number of approximations were previously proposed within the SPT2 approach.
Among these approximations, the SPT2b1 has been considered as the most successful
and accurate one in a large range of fluid densities and for different matrix parameters.
However, at high densities, it can lack accuracy, since it does not take into account
that the maximum packing fraction of a HS fluid in a matrix is limited, not by the geometrical porosity of a matrix
$\phi_{0}$ and the probe particle porosity $\phi$, but by another type of porosity $\phi^{*}$ introduced in our previous studies.
The porosity $\phi^{*}$ is related to the maximal adsorption capacity of a matrix and it
is lower than $\phi_{0}$ and larger than $\phi$. This can be crucial for a fluid in matrices of low porosities and
at high fluid density, especially in the region near close-packing conditions.
Therefore, the approximations SPT2b2 and SPT2b3 taking into account this feature were suggested,
although they still needed a correction because of their poor accuracy. In the present study,
we improved the versions of these approximations, named as SPT2b2$^*$ and SPT2b3$^*$.
We compare these different approximations with the results of computer simulations performed in
the Monte Carlo grand-canonical ensemble. We test the SPT2 approach both for the one- and three-dimensional
cases. We show that the SPT2b3$^*$ provides a very good description of
the chemical potential of a confined fluid, which is better than others. This extends the applicability of
the SPT2 approach to the studies of very dense fluids confined in disordered matrices.
\pacs 61.20.Gy, 61.43.Gy
\end{abstract}

\section{Introduction}

We dedicate this paper to the memory of our good friend, colleague and teacher Jean-Pierre Badiali, who passed away
last year. Among different scientific works by Jean-Pierre, the investigation of phase transition of fluids adsorbed
in deformable porous matrices takes an important place \cite{Vak07jpcb,Vak07,Vak2013}.
In the present paper we consider some improvements of the SPT2 approach which had been developed earlier by us \cite{Pat11} for
the description of thermodynamic properties of the simple hard sphere model of a fluid confined
in a disordered matrix.

Fluids confined in disordered porous materials undergo drastic modifications in their properties, the understanding and prediction of which is highly required for many practical applications \cite{Gelb99}. Starting from the pioneering work of Madden and Gland \cite{Madden88}, different theories have been developed for the description of fluids confined in disordered porous media \cite{Given92,Vlachy1,Vlachy2,Vlachy3,Schmidt05,Rosinberg99,Pizio00}. According to the model of Madden and Glandt, a disordered porous medium is presented as a matrix of quenched configuration of randomly distributed hard spheres.
Despite a comprehensive study of fluids in disordered matrices, the developed approaches were totally numerical.
The first rather accurate analytical expressions for chemical potential and pressure of a hard sphere fluid confined
in a hard sphere (HS) or an overlapping hard sphere (OHS) matrix were obtained only in \cite{Hol09,Pat11,Hol13} by extending the scaled particle theory (SPT) \cite{Reiss59,Reiss60,Leb65}. The developed approach is based on a combination of the exact treatment of a point scaled particle in a hard sphere (HS) fluid with the thermodynamic consideration of a finite size scaled particle.
The exact result for a point scaled  particle in HS fluid confined in a disordered matrix was obtained in \cite{Hol09}. However, the approach proposed in \cite{Hol09} referred to as SPT1 contained a subtle inconsistency appearing when a size of matrix particles is essentially larger than a size of fluid particles. This inconsistency was eliminated in \cite{Pat11} within the framework of a new approach named SPT2. Later on, the SPT2 approach for a hard sphere fluid
in disordered porous media was generalized for one- and two-dimensional cases \cite{Hol10}, and more recently for a fluid of hard convex body particles \cite{Hol14} and for the mixture of hard spheres \cite{Chen} in disordered matrices.

The original expressions derived within the SPT2 approach include two parameters defining the porosity of a matrix.
The first parameter is related to a bare geometry of a matrix. It is the so-called geometrical porosity, $\phi_{0}$, and characterizes the free volume, which is not occupied by matrix particles.
The second one is defined by the chemical potential of a fluid in the limit of infinite dilution,
and it is called a probe particle porosity, $\phi\leqslant\phi_{0}$, characterizing the adsorption of a fluid particle in an empty matrix. It was shown that the SPT2 agrees well with computer simulation data at low fluid densities,
while at medium and at high densities, an essential difference can be obtained.
This difference becomes especially essential when the packing fraction of a fluid $\eta_{1}$ reaches
the values close to the probe particle porosity $\phi$, since the obtained expressions diverge at
$\eta_{1}\rightarrow\phi$.
In order to improve the description in the SPT2 approach, a series of approximations SPT2a, SPT2b, SPT2c and SPT2d
were proposed \cite{Pat11} by a substitution of $\phi$ by $\phi_{0}$ in different places during a derivation
of the expressions for thermodynamic properties of a HS fluid in HS and OHS matrices.
Among these approximations only SPT2a has no divergence at $\eta_{1}=\phi$.
However, it was shown that this approximation systematically underestimates the chemical potential of HS fluid in a matrix even for rather low densities. Other approximations improve the description, but at higher fluid densities, they lead
to an overestimation of the chemical potential due to the divergence at $\eta_{1}=\phi$ as in the original
approximation SPT2. Nevertheless, for that time, the approximation SPT2b was selected as the best one,
since it gave results closest to computer simulation data.
It was noticed that SPT2b is capable of reproducing the simulation results with a good accuracy
at low and medium fluid densities in a wide range of matrix parameters.
On the other hand, the problem of divergence at $\eta_{1}\rightarrow\phi$ still remained.
Thus, to get accurate results at high fluid densities it had to be fixed.

It should be noted that the divergence in the SPT2b approximation is connected only with the logarithmic term $\ln(1-\eta_{1}/\phi)$, which appears in the cases of one and two-dimensional systems as well \cite{Hol10}.
In order to find the way to remove the divergence in the logarithm,
a fluid confined in a disordered matrix was considered in a lower space dimensionality,
namely the one-dimensional case was studied \cite{Hol12}.
By the example of one-dimensional hard rod fluid in a hard rod matrix, the logarithmic term
was modified in such a way as to omit the divergence $\eta_{1}\rightarrow\phi$. Hence, three new approximations of SPT2b
were derived --- SPT2b1, SPT2b2 and SPT2b3. In the last two variants, the new type of porosity parameter $\phi^{*}$ was introduced, and it corresponds to the maximum value of fluid packing fraction of a hard rod fluid in a disordered hard rod matrix. This porosity parameter has a clear physical meaning, since it defines the maximum adsorption capacity of a matrix for a given fluid, and since a matrix is frozen, the porosity $\phi^{*}$ should be less than $\phi_{0}$.

Due to a structure of the expressions obtained in the approximations SPT2b2 and SPT2b3, the divergence appears at $\eta_{1}\rightarrow\phi^{*}$, but in the SPT2b1 it is located at $\eta_{1}\rightarrow\phi_{0}$.
Since $\phi\leqslant\phi^{*}\leqslant\phi_{0}$, the divergence $\eta_{1}\rightarrow\phi^{*}$ is in the correct range of densities. However, some disagreement between the theory and computer simulation data was still observed \cite{Hol12}.
This was also noticed after an application of this concept to the case of three-dimensional system.
Despite the more appropriate physical justification of SPT2b2 and SPT2b3, in general, the variant SPT2b1 appeared
to be better than the latter ones.
Thus, the SPT2b1 has been applied during the recent years
as the most successful one \cite{Hol12, Hol13, Hol15, Kalyuzhnyi14, Hol16, Hol17rpm, Hol17asym, Dong17},
while the SPT2b2 and SPT2b3 needed an improvement. On the other hand, one should bear in mind that SPT2b1 has a wrong asymptotic for highly dense systems (close packing region), for which the divergence should appear at $\phi^{*}<\phi_{0}$.

In this paper, we revise the approximations SPT2b2 and SPT2b3 in order to get an appropriate description
of the thermodynamic properties of a HS fluid in a HS or an OHS matrix at nearly close packing conditions.
For this reason, we propose approximations SPT2b2* and SPT2b3* that are obtained from the Taylor expansion
of logarithmic term around $(\phi-\phi^{*})$. The corresponding expressions for the chemical potential of a one-dimensional system of a hard rod fluid in a hard rod matrix are obtained.
 We also apply the same concept to the case of three-dimensional system of a hard sphere fluid in hard sphere
and overlapping hard sphere matrices.
A comparison with computer simulations has shown that both the approximations SPT2b2$^{*}$ and SPT2b3$^{*}$ give an essential improvement of SPT2b2 and SPT2b3, respectively.

The paper is arranged as follows. The theoretical part with a review of our previous results and a formulation of new approximations is presented in section~\ref{sec2}. Computer simulation details are described in section~\ref{sec3}.
In section~\ref{sec4}, a comparison of different approximations obtained from the SPT2 approach with the computer simulation results is shown and discussed. And finally, we draw some conclusions in the last section.

\section{Theory}\label{sec2}

In this theoretical part, we start from the generalization of our previous results
\cite{Hol09,Pat11,Hol13,Hol10,Hol12} of the application of SPT2 theory and its different modifications for $n$-dimensional hard sphere (HS) fluid in disordered matrices. The key point of the SPT theory consists in deriving
the excess chemical potential of an additional scale particle of a variable size inserted in a fluid. This excess chemical potential is equal to a work needed to create a cavity in a fluid. For a small scaled particle inserted into a HS fluid in the presence of a matrix, the expression for the excess chemical potential is equal to \cite{Hol09}
\begin{equation}
\beta\mu_{\text s}^{\text{ex}}=\ln p_{0}(\lambda_{\text s})-\ln\left(1-\eta_{1}\frac{(1+\lambda_{\text s})^{n}}{p_{0}(\lambda_{\text s})}\right),
\label{holeq2-1}
\end{equation}
where $\beta=1/(k_{\text B}T)$, $k_{\text B}$ is the Boltzmann constant, $T$ is the temperature, $\lambda_{\text s}=R_{\text s}/R_{1}$ is the scaling parameter, $R_{\text s}$ is the radius of a scaled particle, $R_{1}=\sigma_{1}/2$ is the radius of a fluid particle, $n$~is the dimension of space, $\eta_{1}=\rho_{1}v_{1}$ is the fluid packing fraction, $\rho_{1}$ is the fluid density, $v_{1}$ is the volume of a fluid particle. In the one-dimensional case ($n=1$) $v_{1}=2R_{1}$, in the two-dimensional case ($n=2$) $v_{1}=\piup R_{1}^{2}$ and in the three-dimensional case ($n=3$) $v_{1}=\frac43\piup R_{1}^{3}$.
We use here conventional notations \cite{Given92,Schmidt05,Rosinberg99,Pizio00,Hol09,Pat11,Hol13}, where index ``1'' is used to denote a fluid component, the index ``0'' denotes matrix particles, while for the scaled particles the index ``s'' is used.

The term $p_{0}(\lambda_{\text s})=\exp[-\beta\mu_{\text s}^{0}(\lambda_{\text s})]$ is defined by the excess chemical potential $\mu_{\text s}^{0}(\lambda_{\text s})$ of the scaled particle confined in an empty matrix. It has also the meaning of probability to find a cavity created by the scaled particle confined in the matrix in the absence of fluid particles.
It is worth noting that in our study we consider two types of matrices, namely a HS matrix and an OHS matrix.
Both the matrices are presented as quenched configurations of disordered hard spheres.
The difference between them is that the HS matrix is composed of an equilibrated one-component system
of HS particles, while the OHS matrix is a totally random configuration of non-interacting particles (ideal gas configuration).
Therefore, the particles of OHS matrix can freely overlap and compose a porous structure, which fundamentally differs 
from that in the HS matrix.
The $p_{0}(\lambda_{\text s})$ is connected with two types of matrix porosity.
The first one is the geometrical porosity:
\begin{equation}
\phi_{0}=p_{0}(\lambda_{\text s}=0),
\label{holeq2-2}
\end{equation}
which defines the volume of a void between matrix particles. The second one is the so-called probe particle porosity
\begin{equation}
\phi=p_{0}(\lambda_{\text s}=1)=\exp(-\beta\mu_{1}^{0}),
\label{holeq2-3}
\end{equation}
which can be determined by the excess chemical potential of a fluid in infinite dilution $\mu_{1}^{0}$.
In contrast to the geometrical porosity $\phi_{0}$, it also depends on the nature of a fluid.

From (\ref{holeq2-2}) and (\ref{holeq2-3}), the geometrical porosity is obtained as follows:
\begin{equation}
\phi_{0}=1-\eta_{0}
\label{holeq2-4}
\end{equation}
for a HS matrix and
\begin{equation}
\phi_{0}=\exp(-\eta_{0})
\label{holeq2-5}
\end{equation}
for an OHS matrix, where $\eta_{0}=\rho_{0}v_{0}$ is the packing fraction of matrix, $\rho_{0}=N_{0}/V$ is the
matrix density, $N_{0}$ is the number of matrix particles and $V$ is the total volume of a system.
The volume of a single matrix particle is defined as $v_{0}=\sigma_0$ for $n=1$, $v_{0}=\piup\sigma_{0}^{2}/4$ for $n=2$ and
$v_{0}=\piup\sigma_{0}^{3}/6$ for $n=3$, where $\sigma_0=2R_{0}$ is the diameter of the matrix particle.

Using the probe particle porosity given by equation (\ref{holeq2-3}) and
the conventional SPT theory \cite{Reiss59,Reiss60,Leb65} for the pressure $P_0$ of a bulk HS fluid
and for the chemical potential of a fluid in the limit of infinite dilution $\mu_{1}^{0}$, the following expression
in the case of a HS matrix can be derived
\begin{equation}
\phi=(1-\eta_{0})\exp\left(-\frac{\beta P_{0}\eta_{0}}{\rho_{0}}\tau\right), \qquad \frac{\beta P_{0}}{\rho_{0}}=\frac{1}{1-\eta_{0}}
\label{holeq2-6}
\end{equation}
for the one-dimensional system,
\begin{equation}
\phi=(1-\eta_{0})\exp\left[-\left(\frac{2\eta_{0}}{1-\eta_{0}}+\frac{\beta P_{0}\eta_{0}}{\rho_{0}}\tau^{2}\right)\right], \qquad \frac{\beta P_{0}}{\rho_{0}}=\frac{1}{(1-\eta_{0})^{2}}
\label{holeq2-7}
\end{equation}
for the two-dimensional system and
\begin{align}
\phi=(1-\eta_{0})\exp\left\{-\left[\frac{3\eta_{0}\tau}{1-\eta_{0}}+\frac{3\eta_{0}(1+1/2\eta_{0})\tau^{2}}{(1-\eta_{0})^{2}}+\frac{\beta P_{0}\eta_{0}}{\rho_{0}}\tau^{3}\right]\right\}, \qquad
 \frac{\beta P_{0}}{\rho_{0}}=\frac{1+\eta_{0}+\eta_{0}^{2}}{(1-\eta_{0})^{3}}
\label{holeq2-8}
\end{align}
for the three-dimensional system, where $\tau=R_{1}/R_{0}$.
In the case of OHS matrix, one derives from (\ref{holeq2-3}) the following expression for
the probe particle porosity:
\begin{equation}
\phi=\exp[-\eta_{0}(1+\tau)^{n}].
\label{holeq2-8n}
\end{equation}

For a large scaled particle $(R_{\text s}\gg0)$, the excess chemical potential is presented by a thermodynamic expression
for the work needed to create a macroscopic cavity inside a fluid and it can be written in the following form
\begin{equation}
\beta\mu_{\text s}^{\text{ex}}(\lambda_{\text s})=\beta\mu_{\text s}^{0}+\omega(\lambda_{\text s})+\beta\frac{P v_{\text s}}{p_{0}(\lambda_{\text s})}\,,
\label{holeq2-9}
\end{equation}
where $P$ is the pressure of the fluid, $v_{\text s}$ is the volume of a scaled particle. The multiplier $1/p_{0}(\lambda_{\text s})$ appears due to an excluded volume occupied by matrix particles.

According to the ansatz of SPT theory \cite{Pat11,Hol13}, $\omega(\lambda_{\text s})$ can be presented in the form of Taylor expansion, which is cut off at $(n-1)$-th term in $n$-dimensional case. Therefore, in the one-dimensional case one gets
\begin{equation}
\omega(\lambda_{\text s})=\omega_{0}\,,
\label{holeq2-10}
\end{equation}
in the two-dimensional case
\begin{equation}
\omega(\lambda_{\text s})=\omega_{0}+\omega_{1}\lambda_{\text s}\,,
\label{holeq2-11}
\end{equation}
and in three dimensions
\begin{equation}
\omega(\lambda_{\text s})=\omega_{0}+\omega_{1}\lambda_{\text s}+\frac12\omega_{2}\lambda_{\text s}^{2}.
\label{holeq2-12}
\end{equation}
The coefficients $\omega_{0}$, $\omega_{1}$ and $\omega_{2}$ in these expansions can be found from the continuity of $\mu_{\text s}^{\text{ex}}(\lambda_{\text s})$ and its corresponding derivatives $\partial\mu_{\text s}^{\text{ex}}(\lambda_{\text s})/\partial\lambda_{\text s}$ and $\partial^{2}\mu_{\text s}^{\text{ex}}(\lambda_{\text s})/\partial\lambda_{\text s}^{2}$ for $\lambda_{\text s}=0$. After setting $\lambda_{\text s}=1$ the expression~(\ref{holeq2-9}) yields the relation between the pressure $P$ and
the chemical potential $\mu_{1}^{\text{ex}}$ of a fluid confined in a matrix
\begin{equation}
\beta(\mu_{1}^{\text{ex}}-\mu_{1}^{0})=\ln (1-\eta_{1}/\phi_{0})+A\frac{\eta_{1}/\phi_{0}}{1-\eta_{1}/\phi_{0}}+
B\frac{(\eta_{1}/\phi_{0})^{2}}{(1-\eta_{1}/\phi_{0})^{2}}\,,
\label{holeq2-13}
\end{equation}
where
\begin{equation}
A=n-\frac{p'_{0}}{\phi_{0}}+\frac12\left[n(n-1)-2n\frac{p'_{0}}{\phi_{0}}+2\left(\frac{p'_{0}}{\phi_{0}}\right)^{2}-\frac{p''_{0}}{\phi_{0}}\right],
\label{holeq2-14}
\end{equation}
\begin{equation}
B=\frac12\left(n-\frac{p'_{0}}{\phi_{0}}\right)^{2},
\label{holeq2-15}
\end{equation}
$p'_{0}=\partial p_{0}(\lambda_{\text s})/\partial \lambda_{\text s}$ and $p''_{0}=\partial^{2} p_{0}(\lambda_{\text s})/\partial \lambda^{2}_{\text s}$ at $\lambda_{\text s}=0$.

We note that in the one-dimensional case \cite{Hol10} the parameters $A$ and $B$ are
\begin{equation}
A=B=0,
\label{holeq2-16n}
\end{equation}
in the two-dimensional case
\begin{equation}
A=n-\frac{p'_{0}}{\phi_{0}}\,, \qquad B=0,
\label{holeq2-16}
\end{equation}
and for the three-dimensional case $A$ and $B$ are given by (\ref{holeq2-14}) and (\ref{holeq2-15}) when $n=3$.

Using the Gibbs-Duhem equation we obtain the SPT2 result from the relation (\ref{holeq2-13}) for the isothermal compressibility of a fluid:
\begin{align}
\beta\frac{\partial P}{\partial\rho_{1}}&=1+\frac{\eta_{1}/\phi}{1-\eta_{1}/\phi}+(1+A)\frac{\eta_{1}/\phi_{0}}{(1-\eta_{1}/\phi)(1-\eta_{1}/\phi_{0})}
+(A+2B)\frac{(\eta_{1}/\phi_{0})^{2}}{(1-\eta_{1}/\phi)(1-\eta_{1}/\phi_{0})^{2}}
\notag\\
&+2B\frac{(\eta_{1}/\phi_{0})^{3}}{(1-\eta_{1}/\phi)(1-\eta_{1}/\phi_{0})^{3}}\,.
\label{holeq2-17}
\end{align}
A direct integration of this expression leads to the SPT2 result for the excess chemical potential and for the pressure \cite{Pat11}
\begin{align}
\beta (\mu_{1}^{\text{ex}}-\mu_{1}^{0})^\textrm{SPT2}&=-\ln(1-\eta_{1}/\phi)+(A+1)\frac{\phi}{\phi-\phi_{0}}\ln\frac{1-\eta_{1}/\phi}{1-\eta_{1}/\phi_{0}}\nonumber\\
&+(A+2B)\frac{\phi}{\phi-\phi_{0}}\left(\frac{\eta_{1}/\phi_{0}}{1-\eta_{1}/\phi_{0}}-\frac{\phi}{\phi-\phi_{0}}\ln\frac{1-\eta_{1}/\phi}
{1-\eta_{1}/\phi_{0}}\right)
\nonumber\\
&+2B\frac{\phi}{\phi-\phi_{0}}\left[\frac12\frac{(\eta_{1}/\phi_{0})^{2}}{(1-\eta_{1}/\phi_{0})^{2}}-\frac{\phi}{\phi-\phi_{0}}
\frac{\eta_{1}/\phi_{0}}{1-\eta_{1}/\phi_{0}}
+\frac{\phi^{2}}{(\phi-\phi_{0})^{2}}
\ln\frac{1-\eta_{1}/\phi} {1-\eta_{1}/\phi_{0}}\right],
\label{hol2.13}
\end{align}
%
\begin{align}
\left(\frac{\beta P}{\rho_{1}}\right)^\textrm{SPT2}&=-\frac{\phi}{\eta_{1}}\ln\frac{1-\eta_{1}/\phi}{1-\eta_{1}/\phi_{0}}+
(1+A)\frac{\phi}{\eta_{1}}\frac{\phi}{\phi-\phi_{0}}\ln\frac{1-\eta_{1}/\phi}{1-\eta_{1}/\phi_{0}}\nonumber\\
&+(A+2B)\frac{\phi}{\phi-\phi_{0}}\left(\frac{1}{1-\eta_{1}/\phi_{0}}-\frac{\phi}{\eta_{1}}\frac{\phi}{\phi-\phi_{0}} \ln\frac{1-\eta_{1}/\phi}
{1-\eta_{1}/\phi_{0}}\right)\nonumber\\
&+2B\frac{\phi}{\phi-\phi_{0}}\left[\frac12\frac{\eta_{1}/\phi_{0}}{(1-\eta_{1}/\phi_{0})^{2}}-\frac{2\phi-\phi_{0}}{\phi-\phi_{0}}\frac{1}{1-\eta_{1
}/\phi_{0}}+
\frac{\phi}{\eta_{1}}\frac{\phi^{2}}{(\phi-\phi_{0})^{2}}\ln\frac{1-\eta_{1}/\phi}{1-\eta_{1}/\phi_{0}}\right].
\label{hol2.14}
\end{align}
%

As one can see the expressions (\ref{hol2.13})--(\ref{hol2.14}) have two divergences, which appear at $\eta_{1}=\phi$ and $\eta_{1}=\phi_{0}$. Since $\phi<\phi_{0}$, the first divergence at $\eta_{1}=\phi$ occurs at densities lower than the second one. This divergence can strongly worsen the prediction of thermodynamic properties,
especially when the difference between $\phi$ and $\phi_{0}$ increases. In order to improve the SPT2 approach,
different approximations were developed \cite{Pat11}. The most successful among them is the SPT2b approach,
which was derived by replacing $\phi$ with $\phi_{0}$ everywhere in (\ref{holeq2-17}) except the first term.
Such a correction leads to the following modification of the expression (\ref{hol2.13}) for the excess chemical potential of a confined fluid:
\begin{align}
\beta (\mu_{1}^{\text{ex}}-\mu_{1}^{0})^{\text{SPT2b}}&=-\ln(1-\eta_{1}/\phi)+(1+A)\frac{\eta_{1}/\phi_{0}}{1-\eta_{1}/\phi_{0}}+\frac12
(A+2B)\frac{(\eta_{1}/\phi_0)^{2}}{(1-\eta_{1}/\phi_{0})^{2}}
\nonumber\\
&+\frac{2}{3}B\frac{(\eta_{1}/\phi_{0})^{3}}{(1-\eta_{1}/\phi_{0})^{3}}\,.
\label{hol2.15}
\end{align}
We do not present here the expression for the pressure, but it can be found in \cite{Pat11}.
%
This expression for the excess chemical potential essentially improves the description of thermodynamic properties at low and medium densities. However, the divergence at $\eta_{1}=\phi$ remains in the first logarithmic term.
In a study of a hard rod fluid (one-dimensional analog of HS fluid) \cite{Hol12} this term
was presented in the form
\begin{equation}
\ln\left(1-\frac{\eta_{1}}{\phi}\right)=\ln\left(1-\frac{\eta}{\phi_{0}}\right)+
\ln\left[1-\left(\frac{\eta_{1}}{\phi}-\frac{\eta_{1}}{\phi_{0}}\right)\left(1-\frac{\eta_{1}}{\phi_{0}}\right)^{-1}\right].
\label{holeq2-22}
\end{equation}
After the expansion of the last logarithmic term in (\ref{holeq2-22}) it can be rewritten as follows:
\begin{equation}
-\ln\left(1-\eta_{1}/\phi\right)\approx -\ln\left(1-\eta_{1}/\phi_{0}\right)+\frac{\eta_{1}(\phi_{0}-\phi)}{\phi_{0}\phi(1-\eta_{1}/\phi_{0})}\,.
\label{holeq2-23}
\end{equation}
As a result, a new form of the excess chemical potential of a confined fluid is obtained,
and it is referred to as the SPT2b1 approximation:
%
\begin{align}
\beta (\mu_{1}^{\text{ex}}-\mu_{1}^{0})^\textrm{SPT2b1}&=-\ln(1-\eta_{1}/\phi_{0})
+(1+A)\frac{\eta_{1}/\phi_{0}}{1-\eta_{1}/\phi_{0}}+\frac{\eta_{1}(\phi_{0}-\phi)}{\phi_{0}\phi(1-\eta_{1}/\phi_{0})}
\nonumber\\
&+\frac12(A+2B)\frac{(\eta_{1}/\phi_{0})^{2}}{(1-\eta_{1}/\phi_{0})^{2}}+\frac{2}{3}B\frac{(\eta_{1}/\phi_{0})^{3}}
{(1-\eta_{1}/\phi_{0})^{3}}\,.
\label{hol2.18}
\end{align}

Two other approximations called SPT2b2 and SPT2b3 contain the third type of porosity $\phi^{*}$ defined by the maximum value of packing fraction of a fluid in a random porous media.
In order to introduce $\phi^{*}$ in the expression for the chemical potential (\ref{hol2.15}) within the SPT2b2 approximation, the logarithmic term is modified in the following way \cite{Hol12}
\begin{equation}
-\ln \left(1-\eta_{1}/\phi\right)\approx-\ln\left(1-\eta_{1}/\phi^{*}\right)+\frac{\eta_{1}(\phi^{*}-\phi)}{\phi^{*}\phi\left(1-\eta_{1}/\phi^{*}\right)}\,.
\label{holeq2-25}
\end{equation}
Consequently, the chemical potential in the SPT2b2 approximation is derived as
\begin{align}
\beta(\mu_{1}^{\text{ex}}-\mu_{1}^{0})^\textrm{SPT2b2}&=-\ln(1-\eta_{1}/\phi^{*})+\frac{\eta_{1}/\phi_{0}}{1-\eta_{1}/\phi_{0}}(1+A)+\frac{\eta_{1}(\phi^{*}-\phi)}{\phi^{*}\phi(1-\eta_{1}/\phi^{*})}\nonumber\\
&+\frac12(A+2B)\frac{(\eta_{1}/\phi_{0})^{2}}{(1-\eta_{1}/\phi_{0})^{2}}
+\frac23 B\frac{(\eta_{1}/\phi_{0})^{3}}{(1-\eta_{1}/\phi_{0})^{3}}\,.
\label{holeq2-26}
\end{align}

The SPT2b3 approximation can be obtained similar to the SPT2b2 approximation through an expansion of the logarithmic term in the expression (\ref{holeq2-25}) for the chemical potential. As a result, one derives %
\begin{align}
-\ln(1-\eta_{1}/\phi^{*})\approx-\ln (1-\eta_{1}/\phi_{0})+\frac{\eta_{1}/\phi^{*}}{1-\eta_{1}/\phi_{0}}-\frac{\eta_{1}/\phi_{0}}{1-\eta_{1}/\phi_{0}}\,,
\label{holeq2-28m}
\end{align}
\begin{align}
\beta(\mu_{1}^{\text{ex}}-\mu_{1}^{0})^\textrm{SPT2b3}&=-\ln(1-\eta_{1}/\phi_{0})+\frac{\eta_{1}/\phi^{*}}{1-\eta_{1}/\phi_{0}}
+\frac{\eta_{1}(\phi^{*}-\phi)}{\phi^{*}\phi(1-\eta_{1}/\phi^{*})}+A\frac{\eta_{1}/\phi_{0}}{1-\eta_{1}/\phi_{0}}\nonumber\\
&+\frac12(A+2B)\frac{(\eta_{1}/\phi_{0})^{2}}{(1-\eta_{1}/\phi_{0})^{2}}
+\frac23 B\frac{(\eta_{1}/\phi_{0})^{3}}{(1-\eta_{1}/\phi_{0})^{3}}\,.
\label{holeq2-28}
\end{align}
For the sake of brevity, we omit here the expressions for the pressure in the SPT2b1, SPT2b2 and SPT2b3 approximations,
which can be found in \cite{Hol12,HolSmotPat}.

In the present study we also propose two other ways to modify the expression (\ref{hol2.15}), which are based
on the expansion of the logarithmic term in (\ref{hol2.15}) around $(\phi-\phi^{*})$.
The resulting expressions can be considered as alternative ones to the approximations SPT2b2 and SPT2b3.
Therefore, taking into account that
\begin{align}
-\ln(1-\eta_{1}/\phi)\approx-\ln(1-\eta_{1}/\phi^{*})+\frac{\eta_{1}}{(1-\eta_{1}/\phi^{*})}\frac{(\phi^{*}-\phi)}{{\phi^{*}}{\phi^{*}}}\,,
\label{holeq2-30}
\end{align}
instead of SPT2b2 we obtain the chemical potential and pressure of a HS fluid in a disordered matrix
in the approximation, denoted SPT2b2$^{*}$:
\begin{align}
\beta (\mu_{1}^{\text{ex}}-\mu_{1}^{0})^\textrm{SPT2b2$^{*}$}&=-\ln(1-\eta_{1}/\phi^{*})
+\frac{\eta_{1}(\phi^{*}-\phi)}{\phi^{*}\phi^{*}(1-\eta_{1}/\phi^{*})}+(1+A)\frac{\eta_{1}/\phi_{0}}{1-\eta_{1}/\phi_{0}}\nonumber\\
&+\frac12(A+2B)\frac{(\eta_{1}/\phi_{0})^{2}}{(1-\eta_{1}/\phi_{0})^{2}}
+\frac23 B\frac{(\eta_{1}/\phi_{0})^{3}}{(1-\eta_{1}/\phi_{0})^{3}}\,,
\label{holeq2-31}
\end{align}
\begin{align}
\left(\frac{\beta P}{\rho_{1}}\right)^\textrm{SPT2b2$^{*}$}&=-\frac{\phi^{*}}{\eta_{1}}\ln(1-\eta_{1}/\phi^{*})+
\frac{(\phi^{*}-\phi)}{\phi^{*}}
\left[\ln(1-\eta_{1}/\phi^{*})+\frac{\eta_{1}/\phi^{*}}{1-\eta_{1}/\phi^{*}}\right]+\frac{\phi_{0}}{\eta_{1}}\ln(1-\eta_{1}/\phi_{0})\nonumber\\
&+\frac{1}{1-\eta_{1}/\phi_{0}}+\frac{A}{2}\frac{\eta_{1}/\phi_{0}}{(1-\eta_{1}/\phi_{0})^{2}}+\frac{2}{3}B\frac{(\eta_{1}/\phi_{0})^{2}}{(1-\eta_{1}/\phi_{0})^{3}}\,.
\label{holeq2-32}
\end{align}
In the same manner, we derive the expressions for the chemical potential and pressure in the approximation referred to as SPT2b3$^{*}$ and it can substitute the SPT2b3 approximation developed earlier:
\begin{align}
\beta (\mu_{1}^{\text{ex}}-\mu_{1}^{0})^\textrm{SPT2b3$^{*}$}&=-\ln(1-\eta_{1}/\phi_{0})+\frac{\eta_{1}/\phi^{*}}{1-\eta_{1}/\phi_{0}}
+\frac{\eta_{1}(\phi^{*}-\phi)}{\phi^{*}\phi^{*}(1-{\eta_{1}}/{\phi^{*}})}
+A\frac{\eta_{1}/\phi_{0}}{1-\eta_{1}/\phi_{0}}\nonumber\\
&+\frac12(A+2B)\frac{(\eta_{1}/\phi_{0})^{2}}{(1-\eta_{1}/\phi_{0})^{2}}
+\frac23 B\frac{(\eta_{1}/\phi_{0})^{3}}{(1-\eta_{1}/\phi_{0})^{3}}\,,
\label{holeq2-33}
\end{align}
\begin{align}
\left(\frac{\beta P}{\rho_{1}}\right)^\textrm{SPT2b3$^{*}$}&=\frac{(\phi^{*}-\phi)}{\phi^{*}}
\left[\ln(1-\eta_{1}/\phi^{*})+\frac{\eta_{1}/\phi^{*}}{(1-\eta_{1}/\phi^{*})}\right]+\frac{1}{1-\eta_{1}/\phi_{0}}\nonumber\\
&+\frac{(\phi_{0}-\phi^{*})}{\phi^{*}}
\left[\ln(1-\eta_{1}/\phi_{0})+\frac{\eta_{1}/\phi_{0}}{(1-\eta_{1}/\phi_{0})}\right]
+\frac12 A\frac{(\eta_{1}/\phi_{0})}{(1-\eta_{1}/\phi_{0})^{2}}
\nonumber\\
&+\frac23 B\frac{(\eta_{1}/\phi_{0})^{2}}{(1-\eta_{1}/\phi_{0})^{3}}\,.
\label{holeq2-34}
\end{align}

\section{Computer simulation details}\label{sec3}
The grand-canonical ensemble Monte Carlo (GCMC) simulations \cite{Frenkel} were performed
within this study in order to verify an accuracy of different approximations
of the SPT2 approach in one- and three-dimensions.
In the one-dimensional case, a system is presented as a hard rod (HR) fluid in a hard rod (HR)
or an overlapping hard rod (OHR) matrix. In thee dimensions, a system
was considered as a hard sphere fluid in a hard sphere (HS) matrix or an overlapping hard sphere (OHS) matrix.
In the both cases the systems consisted of two components.
The first component was represented by fixed particles of a matrix and the second one by particles of a fluid
that could move in a space not occupied by the matrix particles.
The interactions between particles are defined by the hard-core pair potentials
of matrix and fluid particles with sizes $\sigma_{0}=2R_{0}$ and $\sigma_1=2R_{1}$, respectively.

Two types of a matrix were used in this study. One is composed of the particles distributed
randomly without overlapping (HR or HS matrix), and another one is built totally randomly
with possible overlapping of matrix particles (OHR or OHS matrix).
In all simulations, a number of matrix particles was equal to $N_0=10000$.
Two sizes of matrix particles were considered, $\sigma_{0}=\sigma_{1}$ and $\sigma_{0}=3\sigma_{1}$
in the one-dimensional case and one size, $\sigma_{0}=\sigma_{1}$, in the case of three dimensions.
Also, for the one-dimensional case a fluid in a matrix of point particles was studied ($\sigma_{0}=0$).
It should be noted that all sizes and lengths in our study are presented in units of $\sigma$, which is equal to a diameter of fluid particles, $\sigma_{1}=\sigma$.

Another important parameter of a matrix is its porosity. For a one-dimensional matrix we fixed
the probe particle porosity equal to $\phi=0.35$. In the three-dimensional case, the geometrical
porosity is set equal to $\phi_{0}=0.843$, which depending on the type of a matrix corresponds
to $\phi=0.173$ for a HS matrix or $\phi=0.255$ for an OHS matrix. The parameters of matrix porosities
($\phi$ or $\phi_{0}$) as well as a number of matrix particles $N_{0}$ were used to calculate a size of simulation box
from the relations (\ref{holeq2-4}), (\ref{holeq2-5}) and (\ref{holeq2-6}), taking into account that $V=N_{0}/\rho_{0}$ and $\rho_{0}=\eta_{0}/v_{0}$,
where $v_{0}=\sigma_{0}$ or $v_{0}=\piup\sigma_{0}^3/6$ is a volume of a single particle of
a HR or a HS matrix, respectively. In our study a cubic simulation box was used with the periodical boundary
conditions.

In GCMC simulations, the chemical potential of a fluid is set, and after equilibration
a corresponding fluid density is obtained. Each simulation run starts from a system consisting
of a matrix configuration prepared preliminary and fluid particles of the number density $\rho_{1}=N_1/V=0.3$
placed randomly into the void between matrix particles.
In the case of one-dimensional systems, a number of steps for the equilibration equal to $6\cdot10^5$
and for the production equal to $2\cdot10^5$ appeared to be sufficient to get reliable results
even for the highest fluid densities. In three dimensions,
$2\cdot10^6$ steps were taken for equilibration and $5\cdot10^5$ steps used for production.
To speed up the simulations, the linked cell list algorithm was applied~\cite{Frenkel}.

It should be noted that the results being obtained from simulations are sensitive to a matrix configuration.
Thus, for each set of the parameters, $8$ different matrix configurations were generated and the obtained results
have been averaged over these matrix realizations. This allowed us to obtain fluid densities with a statistical
error less than $0.5\pct$.

Therefore, from the simulations we obtained numerical relations between the fluid density and the chemical potential, which were compared with the results of the SPT2 approach.

\section{Results and discussions}\label{sec4}
\subsection{Hard rod fluid in random porous media}
We use different approximations based on the SPT2 approach for the description of thermodynamic properties
of a HS fluid confined in disordered matrices. First we start with the application of the presented theory
to the one-dimensional case of a confined fluid, i.e., with the description of a hard rod (HR) fluid
in a hard rod (HR) or an overlapping hard rod (OHR) matrix. The hard rods is the one-dimensional analog of hard spheres model and it should allow us to distinguish a role of the logarithmic
term more precisely. According to (\ref{holeq2-16n}) in the case of $n=1$ the parameters $A=B=0$. Some results obtained in the SPT2 approximation and its modifications like the SPT2b, SPT2b1, SPT2b2 and SPT2b3 approximations were already discussed in our previous paper \cite{Hol12}.
It was noticed that the SPTb1, SPT2b2 and SPT2b3 approximations essentially improve the results for thermodynamic
properties in comparison with STP2b. They provide a rather accurate description up to high fluid densities
both in the one- and three-dimensional cases. However, the SPT2b1 approximation fails at the densities near the
close-packing region since it does not take into account the limitation on the maximum packing fraction of a fluid in a matrix, $\eta^{\text{max}}_{1}$, which is directly related to the maximum adsorption capacity of this matrix.
On the other hand, the approximations SPT2b2 and SPT2b3 do not have such a defect since they contain
the third type of porosity, $\phi^{*}$, which defines $\eta^{\text{max}}_{1}$.
In \cite{Hol12} we proposed a generalized expression for $\phi^{*}$:
\begin{equation}
\phi^{*}=\frac{\phi_{0}\phi}{\phi_{0}-\phi}\ln\left(\phi_{0}/\phi\right),
\label{holeq3-1}
\end{equation}
which totally reproduces the analytical expressions derived from the quenched-annealed density
functional theory (QA DFT) by Reich and Schmidt
for the partition coefficient $K(\eta^{r}_{1}\rightarrow1)$ of a HR fluid in a HR
or OHR matrix (see \cite{Reich04}, equations~(38) and (41)).
According to \cite{Reich04} this partition coefficient corresponds to the maximal possible amount of fluid particles,
which can be loaded into a matrix, i.e., $K(\eta^{r}_{1}\rightarrow1)\equiv\eta^{\text{max}}_{1}=\phi^{*}$. It is shown in \cite{Hol12} that the expression (\ref{holeq3-1}) works well for one-dimensional systems.
However, in three dimensions it cannot be considered as an accurate one due to a sphericity of particles,
which cannot fill the whole space with their bodies, in the same way as hard spheres cannot reach the packing
fraction $\eta_{1}=1$ even in the bulk system. Nevertheless, the expression (\ref{holeq3-1}) takes into account
an important feature related to the frozenness of matrix particles. Since, matrix particles cannot move and
they are not in equilibrium with fluid particles, the total packing fraction of matrix and fluid in the system
is always less than one (i.e., $\eta_{0}+\eta_{1}<1$) in any space dimensionality, including the one-dimensional space. To put it differently, the maximum packing fraction of a fluid confined in a matrix is always less than
the geometrical porosity $\eta^{\text{max}}_{1}\equiv\phi^{*}<\phi_{0}$. This feature starts to play an essential
role at high densities of a confined fluid, especially near close-packing conditions.
Therefore, in order to obtain a correct description of thermodynamic properties of a fluid in disordered matrices
at high densities, the applied approximations should contain the porosity $\phi^{*}$.

\begin{figure}[!b]
\begin{center}
\includegraphics[clip,width=0.49\textwidth,angle=0]{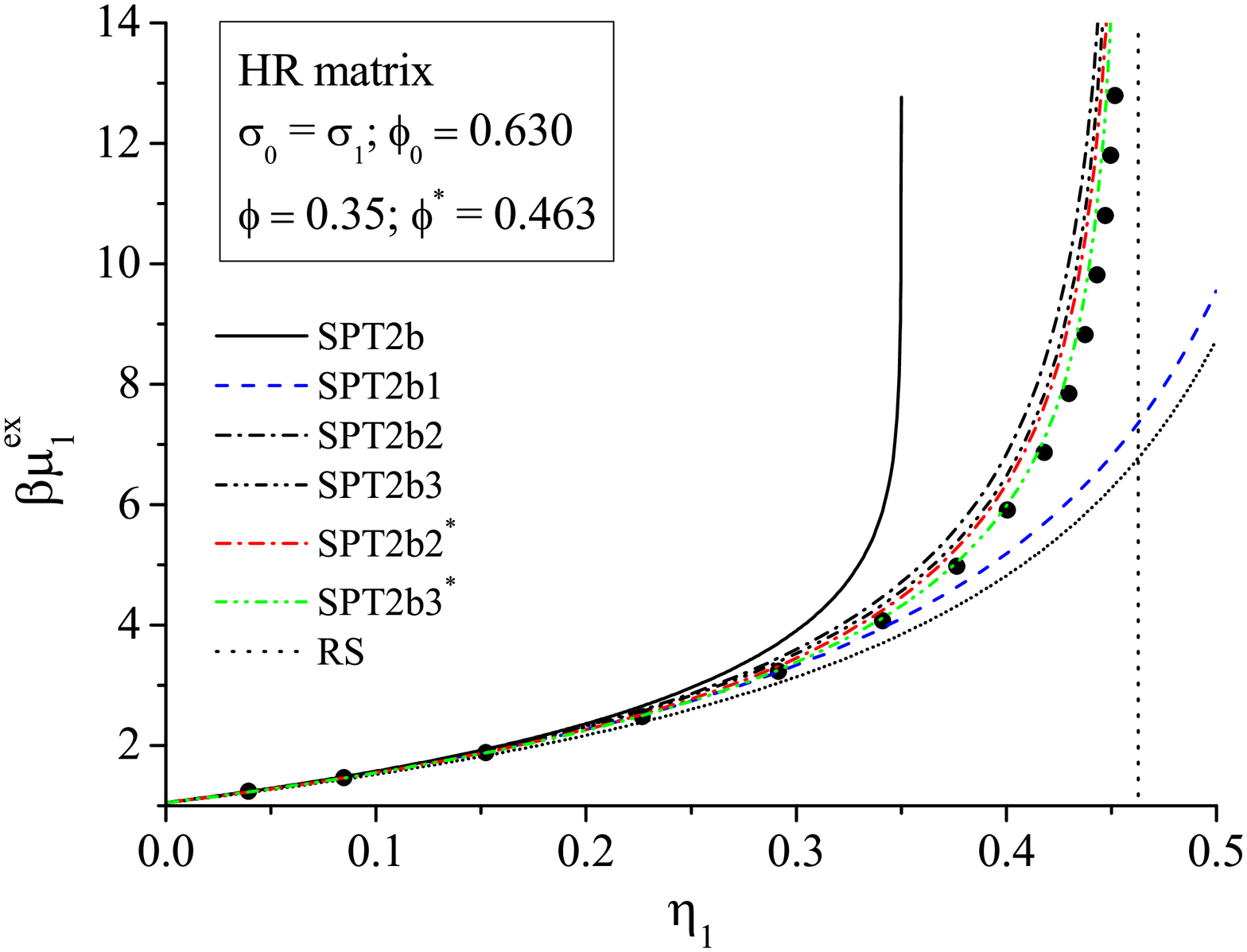}
\includegraphics[clip,width=0.49\textwidth,
angle=0]{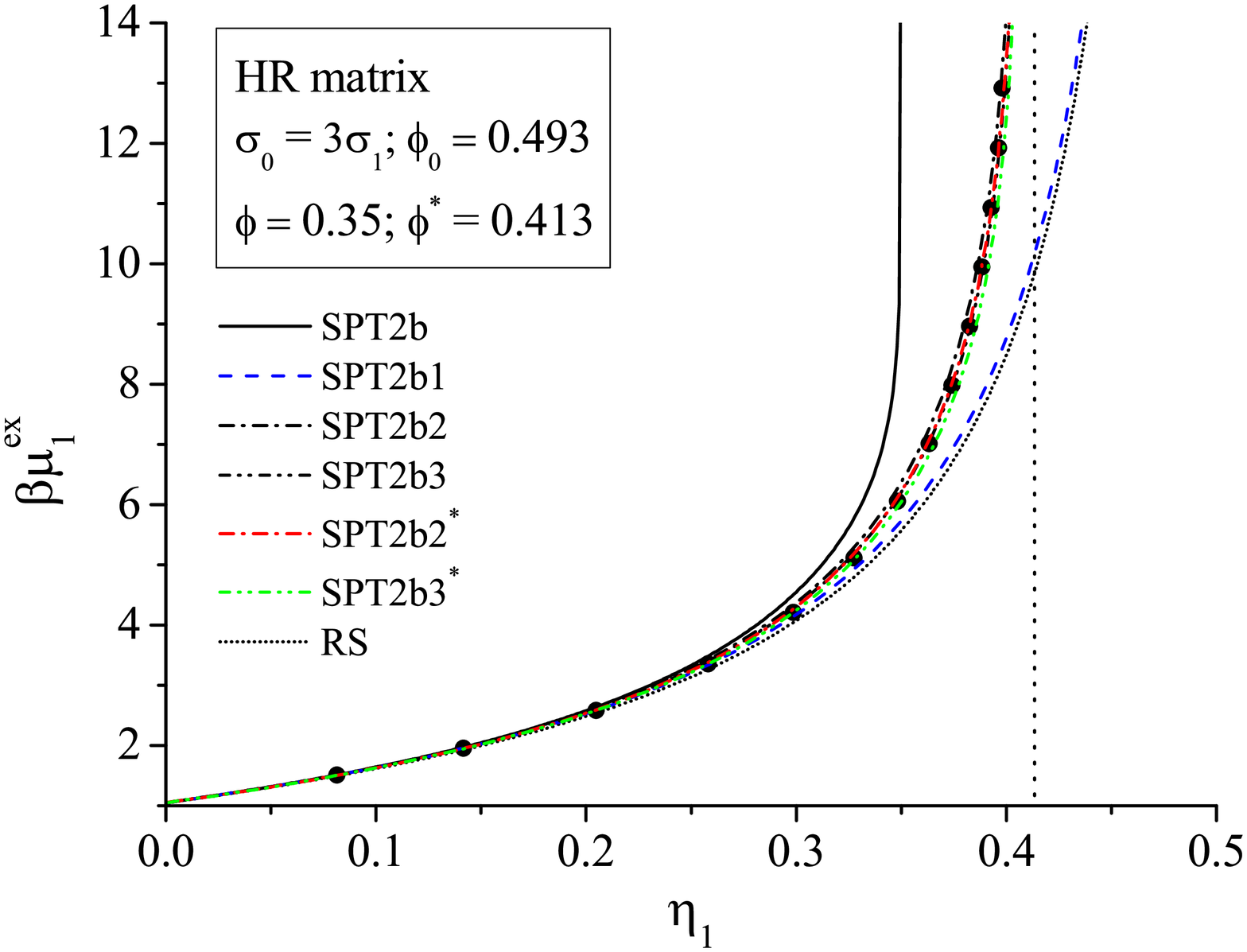} 
\caption{\label{fig:hs1d}
(Color online) The excess chemical potential of a HR fluid in a disordered HR matrix. A comparison of different approximations (lines) with GCMC simulations data (symbols). The vertical dotted line corresponds to the value of maximum packing fraction $\phi^{*}$.
}
\end{center}
\end{figure}
\begin{figure}[!b]
\begin{center}
\includegraphics[clip,width=0.49\textwidth,
angle=0]{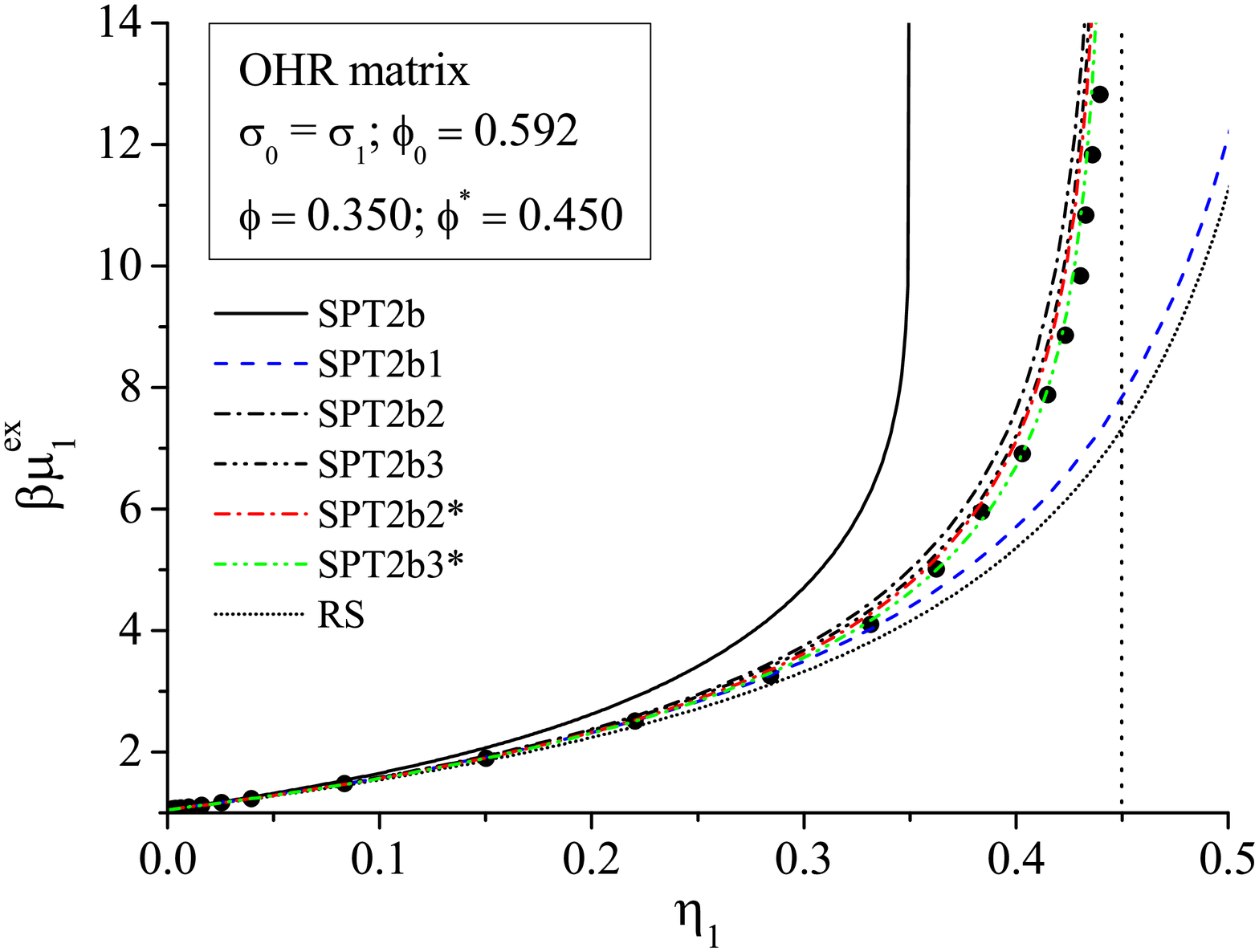}
\includegraphics[clip,width=0.49\textwidth,
angle=0]{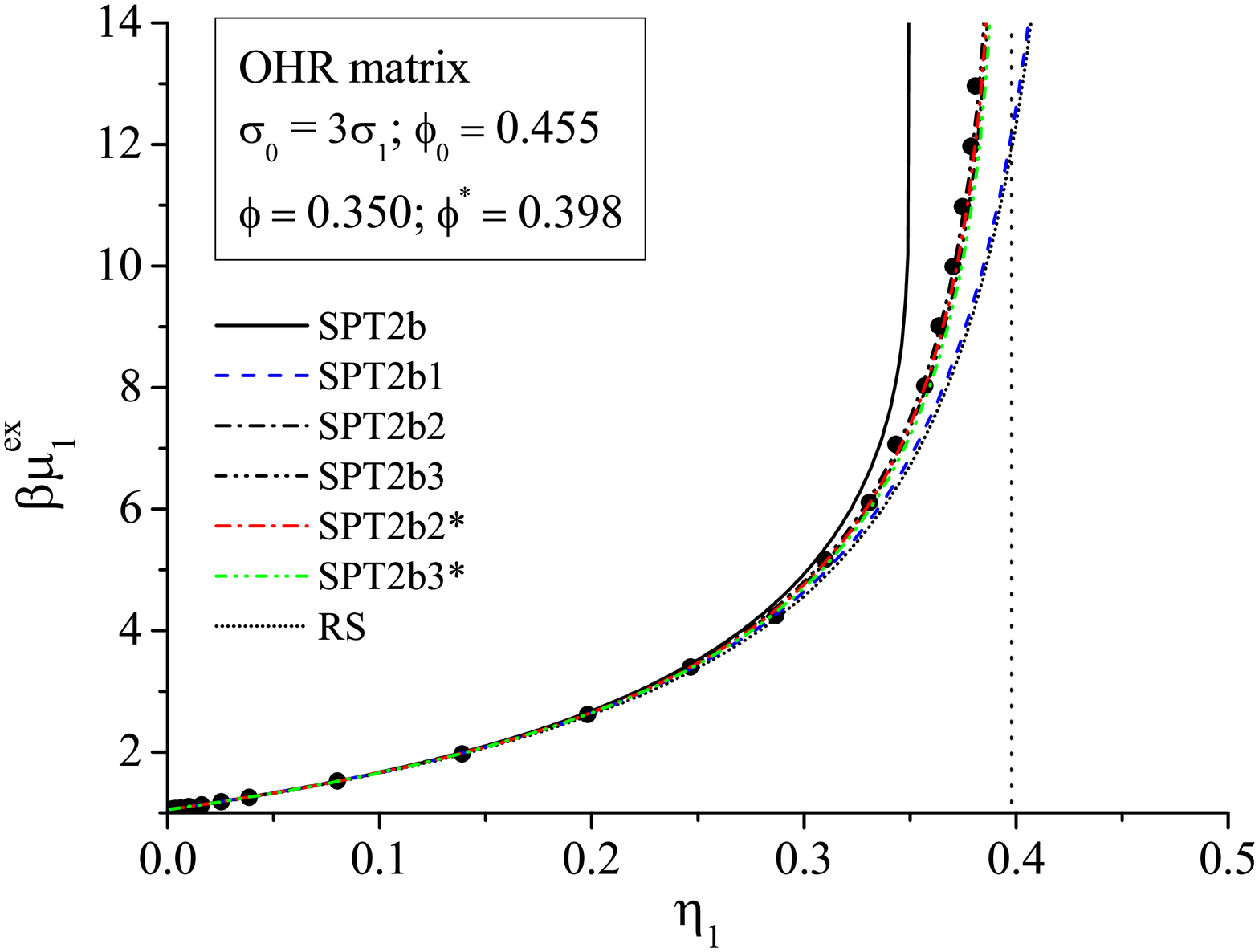} 
\caption{\label{fig:ohs1d}
(Color online) The excess chemical potential of a HR fluid in a disordered OHR matrix. A comparison of different approximations (lines) with GCMC simulations data (symbols). The vertical dotted line corresponds to the value of maximum packing fraction $\phi^{*}$.%
}
\end{center}
\end{figure}

In the present study we introduce two new approximations which take into account $\phi^{*}$,
SPT2b2$^*$ and SPT2b3$^*$, which are aimed at yielding better quantitative results than their
forerunners, the SPT2b2 and SPT2b3.
In order to test an accuracy of all the approximations presented in section~\ref{sec2},
first we consider the one-dimensional system of a HR fluid in a HR or OHR matrix, and
compare the excess chemical potential obtained from the expressions given above
and the computer simulations results taken from our previous paper~\cite{Hol12}.
Similar as it was done in \cite{Hol12} we fix the probe particle porosity $\phi=0.35$ and show
the dependencies of the chemical potential on the packing fraction of a HR fluid in a HR matrix and in an OHR matrix.
We consider two sizes of matrix particles $\sigma_{0}=\sigma_{1}$ (small matrix particles)
and $\sigma_{0}=3\sigma_{1}$ (large matrix particles), which correspond to $\tau=1$ and $\tau=1/3$, respectively.
In figures~\ref{fig:hs1d} and \ref{fig:ohs1d} one can see a comparison of the approximations SPT2b, SPT2b1, SPT2b2, SPT2b3, SPT2b2$^{*}$ and SPT2b3$^{*}$ with the results of GCMC simulations for the excess chemical potential of HR fluid in
disordered HR and OHR matrices. Also, on the same plots we present the results obtained from the analytical
expressions for the chemical potential derived by Reich and Schmidt (RS) for the same systems
with the use of the QA DFT approach (see equations~(42) and (43) in \cite{Reich04}).
We have rewritten these expressions in terms of $\phi_{0}$, $\phi$ and $\tau$ as follows:
\begin{align}
\beta (\mu_{1}^{\text{ex}}-\mu_{1}^{0})^\textrm{RS}&=-\ln\left(1-\frac{\eta_{1}}{\phi_{0}}\right)
+\frac{\tau\eta_{0}q_{0}+\eta_{1}/\phi_{0}}{1-\eta_{1}/\phi_{0}}-\tau\eta_{0}q_{0}\,,
\label{reich-hr}
\end{align}
where the coefficient $q_{0}=1/\phi_{0}$ for a HR matrix and $q_{0}=1$ for an OHR matrix.
As it is seen in figure~\ref{fig:hs1d}~(left-hand panel) the RS formula for a HR fluid in a HR matrix
gives the results close to the SPT2b1 approximation. However, it noticeably underestimates
the chemical potential, as it is shown for the equal sizes of fluid and matrix particles.
In contrast to this, the SPT2b1 perfectly fits the GCMC data, at least up to $\eta_1=0.3$, for the both cases of matrix particle sizes, and then totally fails at the high densities close to $\eta_1=\phi^{*}$ (figure~\ref{fig:hs1d}, left-hand panel).
As it was expected, the original SPT2b approximation leads to overestimation of the chemical potential due to an improper
divergence at $\eta_{1}=\phi=0.35$, thus at $\eta=0.3$, it gives the worst results.
Having a correct location of the divergence at $\eta_{1}=\phi^{*}$, the chemical potentials obtained
in the SPT2b2 and SPT2b3 approximations start to deviate from the GCMC data too early, thus
we get the results in these approximations higher than they should be in the case $\sigma_{0}=\sigma_{1}$.
However, for the size of matrix particles $\sigma_{0}=3\sigma_{1}$, they provide a rather good accuracy, which
is much better than in the SPT2b1. Moreover, for large matrix particles, the SPT2b2 and SPT2b3 almost
coincide with their improved versions, the SPT2b2$^*$ and SPT2b3$^*$ approximations.
However, it is not the case when we have small matrix particles ($\sigma_{0}=\sigma_{1}$), where
the SPT2b2$^*$ and SPT2b3$^*$ are closer to the simulations, and the SPT2b3$^*$ fits the GCMC results
even better than the SPT2b2$^*$ approximation.

For a HR fluid in OHR matrix one can see a very similar picture as it is in the case of a HR matrix.
In figure~\ref{fig:ohs1d}~(right-hand panel) we show a comparison for the same approximations as above. We also present
the results of the RS expression for the chemical potential
of a HR fluid, but in a OHR matrix~(\ref{reich-hr}).
And again for this case we observe an underestimation given by the RS formula in comparison with
the GCMC results. All other conclusions are totally the same as for the HR matrix case.

\begin{figure}[!b]
\begin{center}
\includegraphics[clip,width=0.49\textwidth,angle=0]{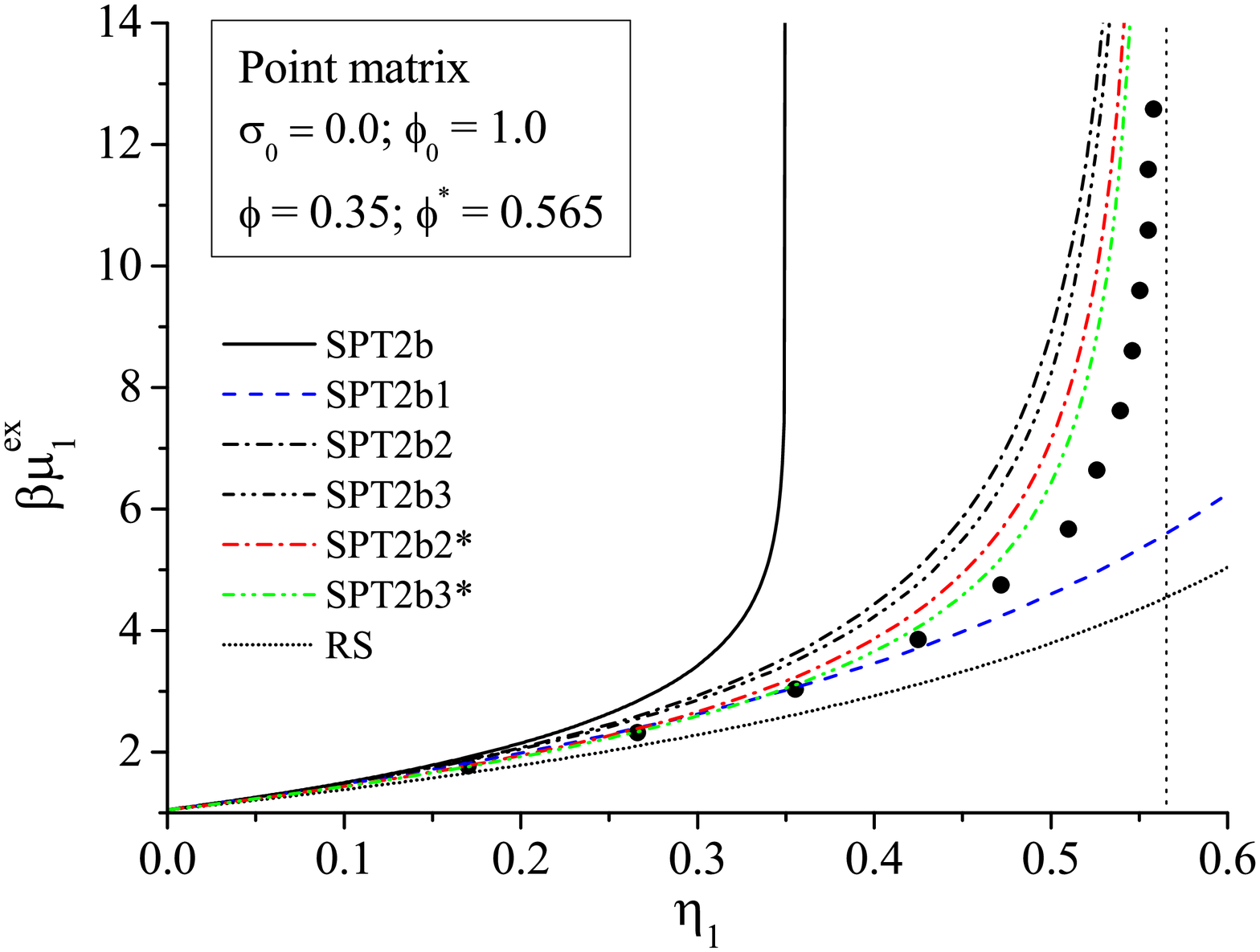} 
\caption{\label{fig:ohs1d0}
(Color online) The excess chemical potential of a HR fluid in a disordered point matrix. A comparison of different approximations (lines) with GCMC simulations data (symbols). The vertical dotted line corresponds to the value of maximum packing fraction $\phi^{*}$.%
}
\end{center}
\end{figure}

It is worth noting that the strongest confinement effect is exhibited for small matrix particles.
Also, the smaller are the sizes of  matrix particles the less accurate results can be predicted by the theory.
On the contrary, with an increase of the size of matrix particles, the prediction of thermodynamic
properties is getting better. Moreover, the best accuracy can be obtained in the limit $\sigma\rightarrow\infty$
(or $\tau\rightarrow0$), if the probe particle porosity $\phi$ is fixed. In this limit all the porosities
become equal and all the approximations presented in this study lead to the same result equivalent
to the bulk-like case with the effective density $\hat{\eta}_1=\eta_1/\phi$.

To make a severe test of an accuracy of the theory by showing its maximum deviation
from simulations, the sizes of matrix particles should be taken as small as possible.
For this purpose, we consider another limit by considering the so-called point matrix,
which is formed by random point particles (the limit $\sigma\rightarrow0$).
In this case, the geometrical porosity tends to one ($\phi_{0}\rightarrow1$),
since the packing fraction of matrix particles tends to zero ($\eta_{0}\rightarrow0$).
At the same time, we fix the probe particle porosity $\phi$, which after taking the limit assumes a finite value
and depends only on the number density of matrix particles $\rho_{0}$ and the size of fluid particles $\sigma_{1}$:
\begin{equation}
\phi=\exp[-\eta_{0}(1+\tau)]\big|_{\sigma_{0}\rightarrow0}=\exp(-\rho_0\sigma_{1}),
\label{holeq3-3}
\end{equation}
where $\eta_{0}=\rho_{0}\sigma_{0}$ and $\tau=\sigma_{1}/\sigma_{0}$.
A set of dependencies of the chemical potential on the packing fraction of a HR fluid confined in a point matrix is presented in figure~\ref{fig:ohs1d0}.
Again, the results obtained in different approximations behave qualitatively in the same way
as in the cases of HR and OHR matrices. However, this time the deviations from the simulations are essentially
larger. This also concerns the SPT2b3$^*$ approximation, which in the previous figures exhibited very accurate results.
Nevertheless, even in this case the SPT2b3$^*$ remains the best among all other approximations. It should also be noted that matrix particles can hardly be smaller than fluid particles. Moreover, they cannot be considered as point ones. However, for methodological reasons, this case is always worth testing.

\begin{figure}[!b]
\begin{center}
\includegraphics[clip,width=0.49\textwidth,
angle=0]{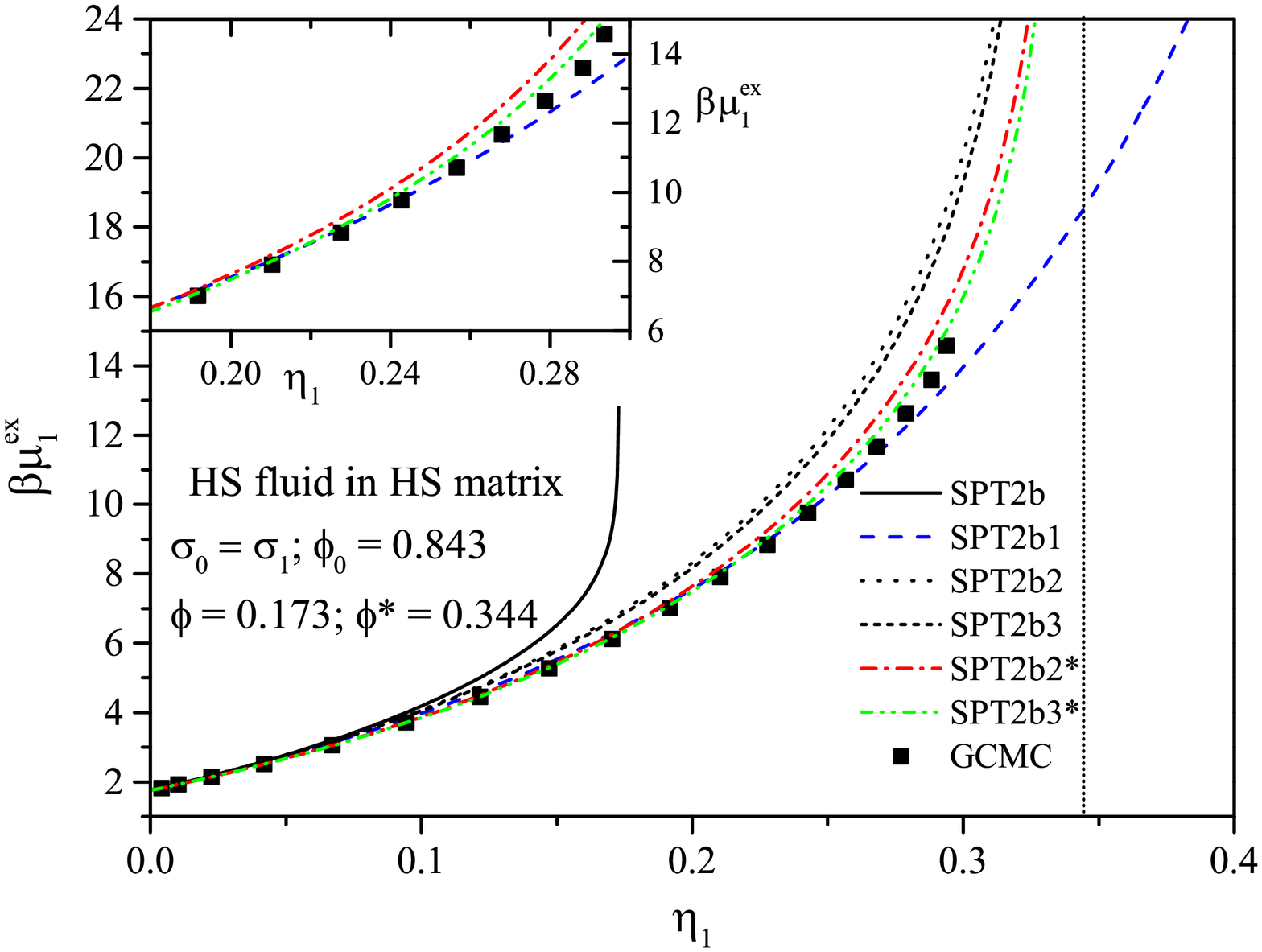}
\includegraphics[clip,width=0.49\textwidth,
angle=0]{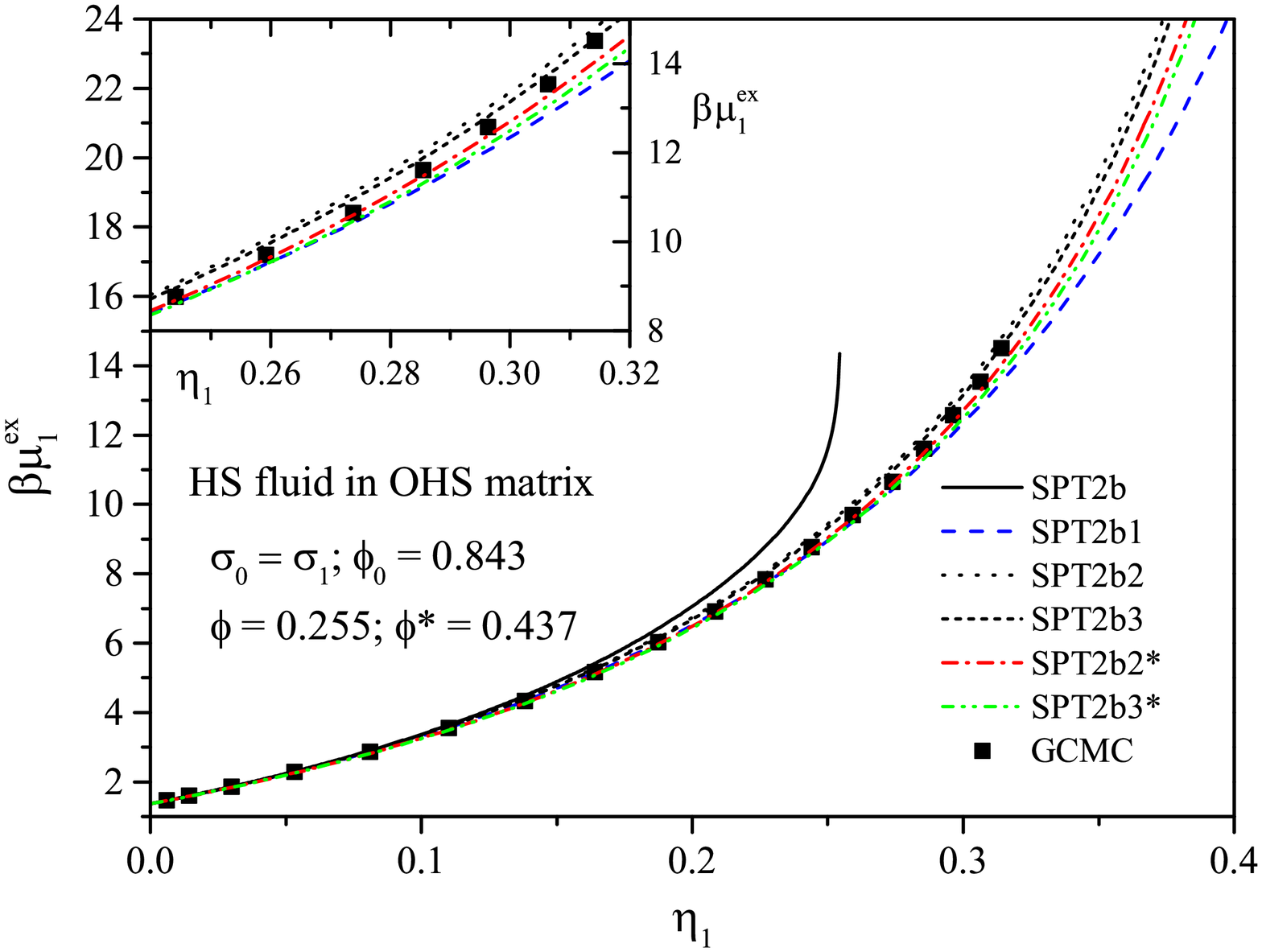} 
\caption{\label{fig:hs3d} (Color online) The excess chemical potential of a HS fluid
in disordered HS and OHS matrices.
A comparison of different approximations (lines) with GCMC simulations data (symbols).
The vertical dotted line corresponds to the value of maximum packing fraction $\phi^{*}$.
}
\end{center}
\end{figure}

\subsection{Hard sphere fluid in random porous media}
From the results obtained for  one-dimensional systems, we conclude that the SPT2b3$^*$ approximation provides
the best and very good accuracy for the chemical potential of a HR fluid confined in disordered HR and OHR matrices
at all fluid densities, including the one quite near the close-packing condition.
However, this is not quite obvious that in the three-dimensional system we will come to the same conclusion.
Therefore, we use the approximations described in this paper for a hard sphere fluid
in disordered hard sphere (HS) and overlapping hard sphere (OHS) matrices.
In this case, the parameters $A$ and $B$ are given by the expressions~(\ref{holeq2-14}) and (\ref{holeq2-15}) when $n=3$. Since the relation (\ref{holeq3-1}) for $\phi^{*}$ is presented in a general form and does not depend on the space dimensionality, we extend its application to the three-dimensional case.
We restrict our calculations to the case of equal sizes of fluid and matrix particles ($\sigma_{0}=\sigma_{1}$ or
$\tau=1$) and to the low value of matrix porosity in order to strengthen the confinement effects on a fluid, and thus clarify all possible drawbacks
of the theory. In figure~\ref{fig:hs3d} we show a comparison of the excess chemical potential obtained
in different approximations with the computer simulations. Two types of matrices (HS and OHS) are taken at
the same geometrical porosity $\phi_0=0.843$, although the probe particle porosities $\phi$ and the maximum adsorption capacity $\phi^{*}$ are different (see the corresponding values in figure~\ref{fig:hs3d}).
Similar to the one-dimensional case, all the approximations correctly reproduce the behaviour of the chemical potential at small fluid densities. At medium fluid densities, the approximation SPT2b has a divergence at $\eta_{1}=\phi$. Thus, it crucially overestimates the chemical potential in this region and does not give any result at higher densities.
The SPT2b1 approximation provides a very good accuracy in a large range of fluid densities up to $\eta_{1}=0.25$,
at which it starts to deviate from the GCMC results, because the SPT2b1 does not take into account the limitation defined by the maximum adsorption capacity of the matrix, i.e., the porosity $\phi^{*}$ is not included
in this approximation. An analysis of the approximations SPT2b2 and SPT2b3 shows that although
they contain the porosity $\phi^{*}$ and have a divergence in the correct place, their accuracy is poor
due to an essential overestimate of the chemical potential, which they provide starting from the medium densities.
This is clearly seen for the case of a HS fluid in a HS matrix (figure~\ref{fig:hs3d}).
However, for a HS fluid in a OHS matrix they are surprisingly good. Moreover, they are as good as the approximations
SPT2b2$^*$ and SPT2b3$^*$, which also have a correct divergence at $\phi^{*}$. We should remind that 
we observed the same picture herein above for a HR fluid in a OHR matrix. Similar to a HR fluid in a HR matrix,
the SPT2b2$^*$ and SPT2b3$^*$ give a much better prediction for the case of a HS fluid in a HS matrix,
and again the SPT2b3$^*$ surpasses SPT2b2$^*$ by accuracy.
At the same time, it is seen that the SPT2b3$^*$ is somewhat worse than SPT2b2$^*$ in the case of OHS matrix
at high densities (figure~\ref{fig:hs3d}). However, the difference in the relative errors of the both approximations
with respect to the simulation results is not so large (within $1.5\pct$ for SPT2b2$^*$ and $2.5\pct$ for SPT2b3$^*$)
as to give an absolute preference to the SPT2b2$^*$. Hence, in general, the SPT2b3$^*$ approximation can be selected as the best one for a description of a chemical potential of a fluid confined in disordered matrices.

\section{Conclusions}
We continue our efforts on improving the SPT2 approach developed for the description of the thermodynamic properties of a hard sphere fluid in disordered matrices \cite{Pat11,Hol13,Hol15}. This approach is based on the scaled particles
theory originally proposed by Reiss, Frisch and Lebowitz \cite{Reiss59} and successfully adapted by us for the case of
quenched-annealed systems like fluids in disordered porous media \cite{Hol09,Pat11,Hol13,Hol15}.
These systems are modelled as a mixture of mobile hard sphere (HS) particles of a fluid immersed in a matrix of frozen
hard spheres. We consider two types of a matrix: (i) a matrix formed by equilibrated hard spheres system (HS matrix);
(ii) matrix particles are randomly distributed (ideal gas-like distribution). Thus, they can overlap and compose the
so-called overlapping hard sphere system (OHS matrix). A void in these two types of matrices forms
qualitatively different porous structures, which are characterized by a specific surface area of pore walls
as well as by their mean curvatures. These factors lead to quantitatively different dependencies of
thermodynamic properties of a confined fluid even at the same matrix porosities\cite{Hol13}.

For the SPT2 approach, a number of approximations have been proposed in our previous studies: SPT2b, SPT2b1, SPT2b2
and SPT2b3. The most accurate was the SPT2b1 approximation, but it did not take into account the fact that due to
the frozenness of matrix particles, the fluid particles cannot be suitably packed into a pore. Thus,
they cannot occupy the whole accessible volume which is defined by the geometrical porosity of matrix, $\phi_{0}$.
Neglecting this fact leads to inappropriate results at high fluid densities, especially near the region of
close-packing conditions.
Therefore, we additionally introduced another porosity parameter, $\phi^{*}$, which is equivalent to
the maximum packing fraction of a fluid in a matrix, $\eta^{\text{max}}_{1}$.
The porosity $\phi^{*}$ is less than $\phi_{0}$ and it corresponds to the maximal adsorption capacity
of a porous medium.
To take this porosity into account, an expression was proposed for it and
the approximations SPT2b2 and SPT2b3 were derived \cite{Hol12}. However, as it was shown herein above, these approximations lack accuracy.

In this paper we considered all the approximations mentioned above and introduced two new approximations, SPT2b2$^*$ and SPT2b3$^*$, which were expected to substitute the SPT2b2 and SPT2b3 approximations as more accurate ones.
We assessed the accuracy of these approximations by comparing them with the GCMC simulations.
Since, the proposed theory can be used for a space of any dimensionality $n\leqslant3$,
first a comparison was made for a one-dimensional case presented as a system of a hard rod (HR) fluid confined
in a hard rod (HR) or an overlapping hard rod (OHR) matrix. Then, we tested the SPT2 approach with different approximations
for three-dimensional systems of a hard sphere (HS) fluid in a hard sphere (HS) or
an overlapping hard sphere (OHS) matrix.
It was shown that the SPT2b2$^*$ and SPT2b3$^*$ approximations are really better than the SPT2b2 and SPT2b3,
and especially this is seen for a HR fluid in a HR matrix and for a HS fluid in a HS matrix.
Between the SPT2b2$^*$ and SPT2b3$^*$ one can choose the SPT2b3$^*$ approximation since in most cases
it provides the best results, although for a HS fluid in an OHS matrix the SPT2b2$^*$ is somewhat better.
It should be also noted that in the presented comparisons, the SPT2b1 approximation showed
a very good accuracy for the description of chemical potential of a HS fluid in HS and OHS matrices
in a wide range of fluid densities. Hence, the SPT2b1 can also be used for the problems where fluid densities are far enough from extreme ones. However, for more general cases, the SPT2b3$^*$ approximation is preferable.

The results obtained within the SPT2 approach can be used as a reference system
for the description of thermodynamic properties of a fluid confined in disordered porous media.
The application of this approach has already been found in many studies of confined liquids
such as simple \cite{Hol13,Hol15} and ionic fluids \cite{Hol16,Hol17rpm,Hol17asym},
network-forming colloids \cite{Kalyuzhnyi14} and binary mixtures \cite{Chen}.
The SPT2 approach has also been developed and used for the description of nematic liquids in matrices \cite{Hol14, HolSmotPat}.
The new approximation SPT2b3$^*$ presented in this paper extends the applicability of the SPT2
due to an essential improvement of its accuracy at high fluid densities.

\section*{Acknowledgements}

MH and TP acknowledge support from the European Union’s Horizon 2020 research and innovation programme under the Marie Sklodowska-Curie (grant No 734276) and the State Fund For Fundamental Research (project N~F73/26-2017).

The computer simulations have been performed on the computing cluster of the Institute for Condensed Matter Physics of NAS of Ukraine~(Lviv, Ukraine).

\ukrainianpart

\title{Удосконалення підходу SPT2 в рамках теорії твердокулькового плину в невпорядкованому пористому середовищі}
\author{М. Головко\refaddr{label1}, Т. Пацаган\refaddr{label1}, В. Донг\refaddr{label2}}
\addresses{
\addr{label1} Інститут фізики конденсованих систем Національної академії наук України,\\
вул. Свєнціцького, 1, 79011 Львів, Україна
\addr{label2} Вища нормальна школа Ліону, хімічна лабораторія, UMR 5182 НЦНД,\\ алея Італії 46, 69364 Ліон, Франція}

\makeukrtitle

\begin{abstract}
\tolerance=3000%

Підхід SPT2 базується на теорії масштабної частинки і розроблений для опису термодинамічних властивостей
твердокулькового (HS) плину в невпорядкованих пористих середовищах.
При використанні цього підходу пористе середовище моделюється як заморожена матриця твердих кульок (HS)
або твердих кульок, які перетинаються (OHS).
Твердокульковий плин, поміщений в матрицю, може рухатися у вільному просторі поміж матричними частинками.
Раніше ряд наближень було запропоновано в рамках підходу SPT2.
Серед цих наближень вважалося, що SPT2b1 є найбільш успішним і точним у широкій області густин
плину і різних параметрів матриці.
Проте, при високих густинах точність може бути недостатньою, оскільки цей підхід не бере до уваги те,
що максимальна упаковка твердокулькового плину в матриці є обмеженою не тільки геометричною пористістю матриці
$\phi_{0}$ і пористістю пробної частинки $\phi$, але й ще одним типом пористості $\phi^{*}$, який був запропонований нами в попередніх дослідженнях.
Пористість $\phi^{*}$ пов'язана із максимальною адсорбційною ємністю матриці, і вона є меншою
ніж $\phi_{0}$ та більшою ніж $\phi$. 
Вона може бути визначальною для плину в матрицях із низькою пористістю та при високих густинах плину,
особливо в області близькій до щільної упаковки.
Тому було запропоновано наближення SPT2b2 і SPT2b3, що враховують цю властивість, хоча й вони 
все ще потребували удосконалення у зв'язку із їх поганою точністю. 
В даному дослідженні ми покращуємо ці наближення і назвали їх SPT2b2$^*$ та SPT2b3$^*$.
Ми порівнюємо ці  наближення із результатами комп'ютерного моделювання, яке проводилося
за допомогою методу Монте-Карло у великому канонічному ансамблі. 
Перевірено підхід SPT2 для одно- і тривимірних випадків. 
Показано, що SPT2b3$^*$ забезпечує дуже добрий опис, який є кращий за інші, для хімічного потенціалу плину в просторовому обмеженні. 
Це розширює область застосування підходу SPT2 на випадок вивчення дуже щільних систем плину в невпорядкованих
матрицях. 
\keywords плини в просторовому обмеженні, пористі матеріали, теорія масштабної частинки, 
твердокульковий плин, хімічний потенціал, пористість
\end{abstract}


\begin{thebibliography}{99}
\bibitem{Vak07jpcb} Vakarin~E.V., Duda Y.,  Badiali~J.P., J. Phys. Chem. B, 2007, \textbf{111}, 2540--2545, \bibdoi{10.1021/jp066460h}.
\bibitem{Vak07} Vakarin~E.V., Dong~W., Badiali~J.P., Physica A, 2007, {\bf 379}, 389--400, \bibdoi{10.1016/j.physa.2006.12.056}.
\bibitem{Vak2013} Vakarin~E.V., Dong~W., Badiali~J.P., Physica A, 2015, {\bf 424}, 294--299, \bibdoi{10.1016/j.physa.2014.12.024}.
\bibitem{Pat11} Patsahan~T., Holovko~M., Dong~W., J. Chem. Phys., 2011, {\bf 134}, 074503, \bibdoi{10.1063/1.3532546}.
\bibitem{Gelb99} Gelb~L.D., Gubbins~K.E., Radhakrishnan~R., Sliwinska-Bartkowiak~M., Rep. Prog. Phys., 1999, {\bf 62}, 1573, \bibdoi{10.1088/0034-4885/62/12/201}.
\bibitem{Madden88} Madden~W.G., Glandt~E.D., J. Stat. Phys., 1988, {\bf 51}, 537, \bibdoi{10.1007/BF01028471}.
\bibitem{Given92} Given~J.A., Stell~G., J. Chem. Phys., 1992, {\bf 97}, 4573, \bibdoi{10.1063/1.463883}.
\bibitem{Vlachy1} Hribar B., Vlachy V., Trokhymchuk A., Pizio O., J. Phys. Chem. B, 1999, \textbf{103}, 5361--5369,\\ \bibdoi{10.1021/jp990253i}.
\bibitem{Vlachy2} Hribar B., Vlachy V., Pizio O., J. Phys. Chem. B, 2000, \textbf{104}, 4479--4488, \bibdoi{10.1021/jp994324p}
\bibitem{Vlachy3} Vlachy V., Dominguez H., Pizio O., J. Phys. Chem. B, 2004, \textbf{108}, 1046--1055, \bibdoi{10.1021/jp035166b}.
\bibitem{Schmidt05} Schmidt~M., J. Phys.: Condens. Matter, 2005, {\bf 17}, S3481, \bibdoi{10.1088/0953-8984/17/45/037}.
\bibitem{Rosinberg99}Rosinberg~M.-L., In: New Approaches to the Problems in Liquid State Theory, NATO Science Series C, Vol.~529, Caccamo~C., Hansen~J.-P., Stell~G. (Eds.), Kluwer, Dordrecht, 1999,  245--278. 
\bibitem{Pizio00}Pizio~O., In: Computational Methods in Surface and Colloidal Science, Surfactant Science Series, Vol.~89, Bor\'owko~M.~(Ed.), Kluwer, Marcell Deker, New York, 2000, 293--346. 
\bibitem{Hol09} Holovko~M., Dong~W., J. Phys. Chem. B, 2009, {\bf 113}, 6360, \bibdoi{10.1021/jp809706n}.
\bibitem{Hol13} Holovko~M., Patsahan~T.,  Dong~W., Pure Appl. Chem., 2012, {\bf 85}, 115, \bibdoi{10.1351/PAC-CON-12-05-06}.
\bibitem{Reiss59} Reiss~H., Frisch~H.L., Lebowitz~J.L., J. Chem. Phys., 1959, {\bf 31}, 369, \bibdoi{10.1063/1.1730361}.
\bibitem{Reiss60} Reiss~H., Frisch~H.L., Helfand~E., Lebowitz~J.L., J. Chem. Phys., 1960, {\bf 32}, 119, \bibdoi{10.1063/1.1700883}.
\bibitem{Leb65}Lebowitz~J.L., Helfand~E., Praestgaard~E., J. Chem. Phys., 1965, {\bf 43}, 774, \bibdoi{10.1063/1.1696842}.
\bibitem{Hol10}Holovko~M.F., Shmotolokha~V.I.,  Dong~W., Condens. Matter Phys., 2010, {\bf 13}, 23607,\\ \bibdoi{10.5488/CMP.13.23607}.
\bibitem{Hol14} Holovko~M., Shmotolokha~V.,  Patsahan~T., J. Mol. Liq., 2014, {\bf 189}, 30, \bibdoi{10.1016/j.molliq.2013.05.030}.
\bibitem{Chen}Chen~W., Zhao~S.L., Holovko~M., Chen~X.S., Dong~W.,  J. Phys. Chem. B, 2016, {\bf 120}, 5491,\\ \bibdoi{10.1021/acs.jpcb.6b02957}.
\bibitem{Hol12}Holovko~M.F., Patsahan~T.M., Dong~W.,  Condens. Matter Phys., 2012, {\bf 15}, 23607, \bibdoi{10.5488/CMP.15.23607}.
\bibitem{Hol15}Holovko~M., Patsahan~T., Shmotolokha~V.,  Condens. Matter Phys., 2015, {\bf 18}, 13607, \bibdoi{10.5488/CMP.18.13607}.
\bibitem{Kalyuzhnyi14} Kalyuzhnyi~Yu.V., Holovko~M., Patsahan~T., Cummings~P.T., J. Phys. Chem. Lett., 2014, {\bf 5}, 4260, \bibdoi{10.1021/jz502135f}.
\bibitem{Hol16} Holovko~M., Patsahan~O., Patsahan~T., J. Phys.: Condens. Matter, 2016, {\bf 28}, 414003,\\ \bibdoi{10.1088/0953-8984/28/41/414003}.
\bibitem{Hol17rpm} Holovko~M., Patsahan~T., Patsahan~O., J. Mol. Liq., 2017, \textbf{228}, 215--223, \bibdoi{10.1016/j.molliq.2016.10.045}.
\bibitem{Hol17asym}	Holovko~M., Patsahan~T., Patsahan~O., J. Mol. Liq., 2017, \textbf{235},  53--59, \bibdoi{10.1016/j.molliq.2016.11.030}.
\bibitem{Dong17} Qiao C.Z., Zhao~S.L., Liu~H.L., Dong~W., J. Chem. Phys., 2017, \textbf{146}, 234504, \bibdoi{10.1063/1.4984773}.
\bibitem{Reich04} Reich~H., Schmidt~M., J. Stat. Phys., 2004, {\bf 116}, 1683, \doi{10.1023/B:JOSS.0000041752.55138.0a}. 
\bibitem{HolSmotPat}Holovko~M., Shmotolokha~V., Patsahan~T., In: Physics of Liquid Matter: Modern Problems, Springer Proceedings in Physics, Vol.~171, Bulavin~L., Lebovka~N.~(Eds.), Springer, Cham, 2015, 3--30,\\ \doi{10.1007/978-3-319-20875-6_1}.
\bibitem{Frenkel} Frenkel~D., Smith~B., Understanding Molecular Simulations, Academic, San Diego, 1995.
\end{thebibliography}
\end{document}